%% file: priyatikanto-ms.tex
\documentclass{aastex}

\usepackage{xcolor,graphicx}
\usepackage{tikz}
\usepackage{multirow}
\usepackage{spr-astr-addons}
\usepackage{url}
\usepackage[colorlinks,citecolor=blue,urlcolor=blue,linkcolor=blue]{hyperref}

\journal{New Astronomy}
\newcommand{\bkdee}{\mbox{BKDE-$e$}}

\begin{document}

\title{The Implementation of Binned Kernel Density Estimation\\ to Determine Open Clusters' Proper Motions:\\ Validation of the Method}
\shorttitle{Implementation of \bkdee}
\shortauthors{Priyatikanto \& Arifyanto}

\author{R. Priyatikanto\altaffilmark{1,2}} \and
\author{M. I. Arifyanto\altaffilmark{1,3}}
\email{rho2m@hotmail.com}

\altaffiltext{1}{Astronomy Department, Institut Teknologi Bandung, Indonesia\\ email: \texttt{rho2m@hotmail.com}}
\altaffiltext{2}{Space Science Centre, National Institute for Aeronautics and Space, Bandung, Indonesia}
\altaffiltext{3}{Astronomy Research Group, Faculty of Mathematics and Natural Sciences, Institut Teknologi Bandung, Indonesia}

\begin{abstract}
Stellar membership determination of an open cluster is an important process to do before further analysis. Basically, there are two classes of membership determination method: parametric and non-parametric. In this study, an alternative of non-parametric method based on \emph{Binned Kernel Density Estimation} that accounts measurements errors (simply called \bkdee) is proposed. This method is applied upon proper motions data to determine cluster's membership kinematically and estimate the average proper motions of the cluster. Monte Carlo simulations show that the average proper motions determination using this proposed method is statistically more accurate than ordinary Kernel Density Estimator (KDE). By including measurement errors in the calculation, the mode location from the resulting density estimate is less sensitive to non-physical or stochastic fluctuation as compared to ordinary KDE that excludes measurement errors. For the typical mean measurement error of 7 mas/yr, {\bkdee} suppresses the potential of miscalculation by a factor of two compared to KDE. With median accuracy of about 93\%, {\bkdee} method has comparable accuracy with respect to parametric method (modified Sanders algorithm). Application to real data from The Fourth USNO CCD Astrograph Catalog (UCAC4), especially to NGC 2682 is also performed. The mode of member stars distribution on Vector Point Diagram is located at $\mu_{\alpha}\cos\delta=-9.94\pm0.85$ mas/yr and $\mu_{\delta}=-4.92\pm0.88$ mas/yr. Although the {\bkdee} performance does not overtake parametric approach, it serves a new view of doing membership analysis, expandable to astrometric and photometric data or even in binary cluster search.
\end{abstract}


\section{Introduction}
Stellar clusters are essential building blocks of galaxies such that comprehensive study upon them brings more complete understanding toward structure and evolution of Galaxy \citep[e.g][]{piskunov06,frinchaboy08,lepine11}. Based on the latest statistics by W. S. Dias in January 2013 there are 2174 catalogued open clusters in Milky Way (\url{http://www.astro.iag.usp.br/~wilton/}) where the average proper motions of 55\% of these open clusters have been determined by several of researchers \citep[e.g.][]{dias02,kharchenko05}. These determinations, of course, need to be done after careful membership determination since the existing field stars obscure those Galactic open clusters.

Special characteristics of open clusters as spatial, kinematic, and photometric agglomeration of stars are the foundation of membership determination that basically distinguish cluster members out off contaminating field stars. Several methods have been developed to evaluate the membership probability of any star inside sampling radius of open cluster. One of the oldest parametric method has been introduced by \citet{vasilevkis58} and developed by several authors \citep{sanders71,cano85,zhao90,zhao2006,krone2010}, employing proper motions data. This method assumes that the proper motions distribution of member and non-member populations follow a normal bivariate distribution function \citep{vasilevkis58}. Two normal distribution function with different parameters are fitted to the data iteratively. The latest development of this method includes measurement errors into calculation and claims that the algorithm may provide intrinsic velocity dispersion for mass calculation \citep{zhao90,zhao2006}. Both in-depth cluster study \citep[e.g.][]{wira06} and batch analysis \citep{dias02p,dias06} employ this method for membership determination.

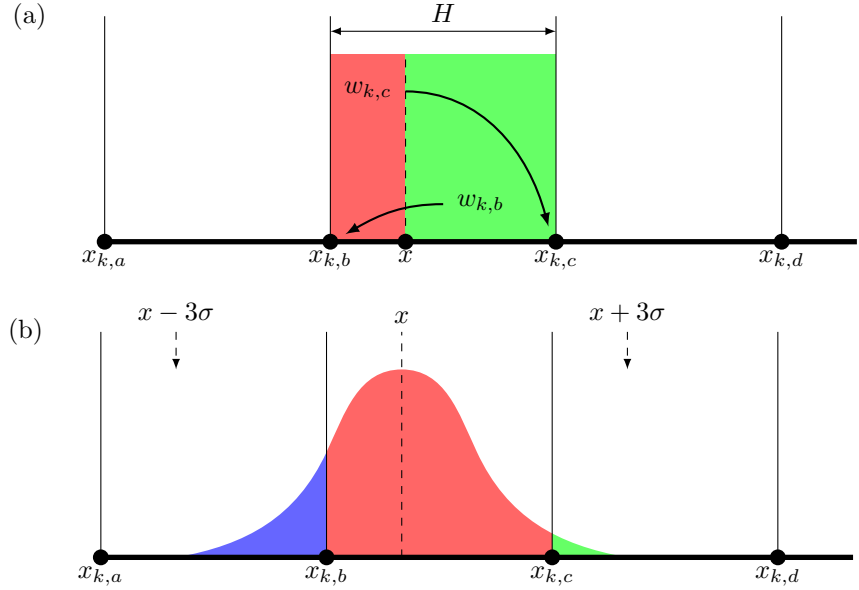
\begin{figure*}[!ht]
\centering
\usetikzlibrary{arrows}
\input{priyatikanto-fig1a.tex}\\
\vskip5pt
\input{priyatikanto-fig1b.tex}
\caption{Illustration of linear binning process of one-dimension data point $x\pm\sigma$ contained by $H$-sized bins. Without error calculation (a), data point $x$ only gives its weight to two neighbouring knots ($x_{k,b}$ and $x_{k,c}$), proportional to its closeness. For example, $x$ gives its partial weight to knot $x_{k,b}$ proportional to the dark-grey area ($w_{k,b}$). While in error calculation (b), that data point gives weight to knot $x_{k,a},\ldots,x_{k,d}$ that confine the whole probability distribution of $x\pm\sigma$.}
\label{fig:binning}
\end{figure*}

The basic assumptions of parametric method face problem when dealing with high-uncertainty data, non-Gaussian distribution of field stars' kinematics, or overlapping distribution between member and non-member stars \citep{cano90}. These conditions drive the emergence of non-parametric methods in membership probability calculation \citep{cano90, galadi98}. Within the framework of this method, the kinematic distribution of both population are constructed using density estimates such as Kernel Density Estimation (KDE, \citet{silverman86}). Non-parametric nature of this approach serves flexibility that enable researcher to find multimodality such as the overlapping cluster NGC 1750 and NGC 1758 \citep{galadi98}.

However, the density estimation process as the hearth of non-parametric approach requires greater computational effort compared to the iteration process of parametric methods. For this case, binning process (namely \emph{binned} KDE) gives alternative and faster way to estimate density distribution without reducing the accuracy \citep{wand94}. This scheme reduces the computational work from $O(n^2)$ or $O(nn_k^2)$ (for direct or ordinary binned KDE) to $O(n+n_k^2)$, which $n$ and $n_k$ represent number of evaluated data points and employed bins. The lower computational effort is essential while dealing with overwhelming survey data coming in the future, e.g. the rise of kinematic data from HST and forthcoming GAIA survey. Faster algorithm also serves advantage in handling high-dimensional membership analysis, e.g. simultaneous membership determination using astrometric, kinematic and photometric data. Besides, \emph{binned} KDE provides a chance to account measurement errors during density estimation.

In this study, the performance of kernel based density estimation for membership determination is explored. A new scheme of BKDE that evaluates measurement errors through numerical technique is proposed. We name it {\bkdee}. Monte Carlo simulations are drawn to test the method and compare it to the parametric method (modified Sanders algorithm, \citet{zhao90}) and the basic KDE method.

This paper is organized as follows: In \autoref{sec:bkde} fundamental concepts of KDE, BKDE, and proposed {\bkdee} that accounts errors are explained. Then, Monte Carlo simulations for validation are presented in \autoref{sec:validation}. The result of these simulations is discussed in \autoref{sec:result}, accompanied by the application to the real kinematic data for NGC 2682 obtained from US Naval Observatory CCD Astrograph Catalog \citep{zacharias13}. At last, the conclusions are given in \autoref{sec:conclusions}.

\section{Non-parametric Membership Analysis}
\label{sec:bkde}
Density distribution of data points can be estimated through various ways. One of the well-known density estimation is kernel based which can be applied to univariate or multivariate data \citep{silverman86}. Let $x=x_1,\ldots,x_n$ and $y=y_1,\ldots,y_n$ be two-dimensional data. Density distribution of these data points with respect to $(x,y)$ can be estimated using the following equation:
\begin{equation}
\hat{f}(x,y)=\dfrac{1}{nh_xh_y}\sum_{i=1}^{n}\sum_{j=1}^{n}K\left(\dfrac{x-x_i}{h_x}\right) K\left(\dfrac{y-y_j}{h_y}\right),
\label{eq:kernel}
\end{equation}
where $K(x')$ is kernel function, while $h_x$ and $h_y$ represent kernel width or smoothing parameter.

There are several kernel function to be employed for evaluating \autoref{eq:kernel}, one of which Gaussian kernel. The choice of function among the most widely used kernel function is not as crucial as the choice of kernel width, $h$. This parameter determines smoothing level and of course the accuracy of the estimate. To minimize errors, \citet[][p 45]{silverman86} proposed simple rule-of-thumb to determine optimal value of $h$, especially for normally distributed data:
\begin{equation}
h_{\text{opt}}=1.06\sigma n^{-0.2},
\end{equation}
where $\sigma$ is standard deviation of the data. This rule work well if the data is normally distributed, but it may oversmooth if the data is multimodal. For the case of equal mixture of standards normal distributions ($\sigma=1$) with means separated by two or more, \citet{silverman86} showed that the rule oversmooth by a factor of two. Fortunately, Solar neighbour open clusters have relatively low proper motions (less than $\sim10$ mas/yr) compared to the field stars' proper motion dispersion (of order 10 mas/yr). This argument implies that separation between members and field stars distributions in vector point diagram is usually less $2\sigma$ and Silverman's rule becomes appropriate.

\subsection{Binned Kernel Density Estimation (BKDE)}
When dealing with large number of data, it is more convenient to pre-bin the data before density estimation \citep[][and reference therein]{wand94}. In this scheme, a number of knots ($x_k,y_k$) are generated as the representation of relevant bins. Each data point has its weight with respect to neighbouring knots. Then kernel density estimate of each knot is multiplied by binning count ($c_{k}$) that represents the total weights ($w_k$) from surrounding data points (see \autoref{fig:binning}). \citet{wand94} stated that linear binning, that assigns weight proportionally to the closeness of any data point and nearby knot, provides better count compared to simple counting rule. Adoption of this scheme may improve ordinary binned KDE as used by \citet{nunez07}.

Density estimate for each knot can be evaluated using the following equation:
\begin{multline}
\hat{f}(x_k,y_k)=\dfrac{1}{nh_xh_y}\sum_{i=1}^{n_{k,x}}\sum_{j=1}^{n_{k,y}}K\left(\dfrac{x_k-x_{k,i}}{h_x}\right)\\ K\left(\dfrac{y_k-y_{k,j}}{h_y}\right)c_{k,ij}.
\end{multline}

Computation effort for linear binning is $O(n)$, while the kernel evaluation is $O(n_k^2)$ such that total computation effort for BKDE scheme is $O(n+n_k^2)$ which is lower than ordinary KDE evaluation. However, binning process may eliminate some informations and generate bias between BKDE and KDE. To minimize this effect, \citet{wand94} gave a complex polinomial rule to determine appropriate number of grid $n_k$, for example  $n_k\geq32$ for 10,000 normally distributed data which is appropriate to open cluster kinematic data \citep[see Table 1 of][]{wand94}. Using this rule, the achieved relative mean integrated error,
\begin{equation}
RMISE = \dfrac{\text{binning error}}{\text{total expected error}}\approx1\%.
\end{equation}

\subsection{Dealing with uncertainty}
Let $x=x_1,\ldots,x_n$ be one-dimensional data with uncertainty of $\sigma=\sigma_1,\ldots,\sigma_2$. Each data point with its uncertainty can be interpreted as normal probability density function (PDF) centered at $x_i$ and having $\sigma_i$ deviation. Instead of giving its weight into nearby knots, each data point spread its weight along the PDF that may include more surrounding knots as illustrated in \autoref{fig:binning}. For this case, total weights form $x_i$ to knot $x_{k,b}$ and $x_{k,c}$ can be calculated using the following equations:
\begin{align}
\label{eq:integral1}
w_{k,b}&=\int_{x_b}^{x_c}f(x')dx'-\dfrac{1}{H}\int_{x_b}^{x_c}(x'-x_b)f(x')dx',
\\
\label{eq:integral2}
w_{k,c}&=\dfrac{1}{H}\int_{x_b}^{x_c}(x'-x_b)f(x')dx',
\end{align}
where $f(x')=f(x'|x,\sigma)$ expresses normal PDF of $x_i$, while $H$ is the bin size. This formulation is also applied for another relevant knots, e.g. $x_{k,a}$ and $x_{k,d}$ in \autoref{fig:binning}. Then, the binning count is merely the total weights obtained by each knot:
\begin{equation}
\label{eq:count}
c_{k,i}=\sum_{j=1}^nw_{k,i}(x_j).
\end{equation}
This $1D$ binning scheme can be expanded for multidimensional cases where \autoref{eq:integral1} and \autoref{eq:integral2} can be evaluated through the following two approaches:
\begin{itemize}
\item\textbf{Analytic approach}: integration of the Gaussian PDF can be done analytically for each binning segment (definite integral). But, if the uncertainty $\sigma$ much greater than the bin size $H$, analytic evaluation of these equations may takes more effort.
\item\textbf{Numerical integration}: in this scheme, each data point $x_i$ generates $m$ random dummy points with normal distribution. Then, every dummy point gives $1/m$ weight to its neighbouring knots. This scheme needs extra computational effort of $O(nm)$ for binning purpose, $m$ times larger than ordinary binning scheme. Beside that, employing random numbers means stochastic variation of the obtained solution. The emerging error follows Poisson statistics that decreases by $\sqrt{m}$.
\end{itemize}

Among these two approaches, we use numerical integration for the following analysis which is considerably more practical.

\section{Monte Carlo Simulation}
\label{sec:validation}

To examine the behaviour and performance of the proposed method, 1000 Monte Carlo simulations are drawn. In this simulation, synthetic data that consist of positions, magnitude and proper motions with errors are created to mimic the real data. Both member and non-member populations are created.

\subsection{Synthetic Data}
\begin{figure}[ht]
\centering
\includegraphics[width=0.48\textwidth]{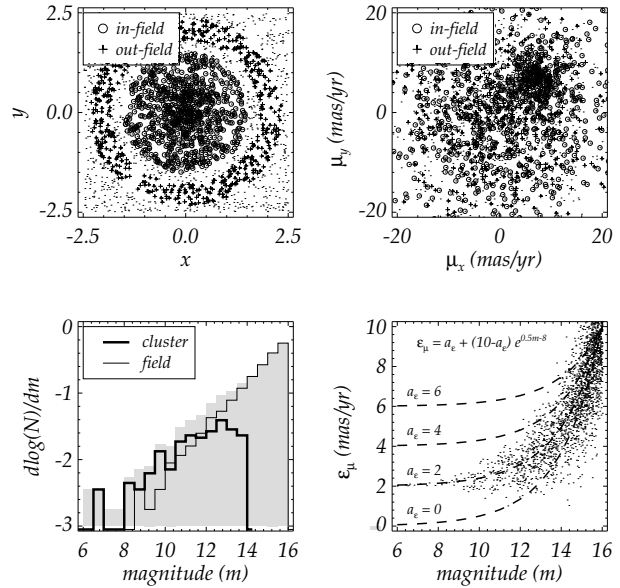}
\caption{Characteristics of the generated synthetic data for Monte Carlo simulations. Upper-left: positional data; upper-right: proper motions data; lower-left: luminosity function; lower-right: proper motions error as function of magnitude.}
\label{fig:monte}
\end{figure}

Spatial distribution of member stars in the observational plane is assumed to follow Gaussian distribution centered at $(0,0)$ with deviation of $\sigma=0.4$ units. By this definition, more than 99\% of member stars are located inside the radius of $r=1.2$ units around cluster center. This radius is analogous to the tidal radius of real cluster which usually determined from the fitting of appropriate density profile. In practice, this radius could be determined non-parametrically using positional data, e.g. the radius where the density estimate drops to the density of the field stars. On the other hand, non-member stars are distributed uniformly within square of $5\time5$ units around cluster center. The adopted unit is arbitrary since it does not have physical meaning toward further analysis. Positional data is used to select \emph{in-field} stars which are considered as main sample ($r<1.50$) and control sample of \emph{out-field} stars ($1.80\leq r\leq2.34$). Definition of these limiting radii is analogous to the \emph{annulus} and \emph{dannulus} in aperture photometry analysis. The innermost/sampling radius is chosen to encompass member stars as complete as possible and simultaneously minimize the contamination from field stars \citep{sanchez09}. Inner and outer radii for out-field selection are carefully chosen such that the out-field ring has similar area to the in-field and covers kinematically similar field stars compared to field stars inside the in-field area.

Magnitude distribution of field stars follow power law distribution function as observed in USNO CCD Astrograph Catalog  \citep{zacharias13}. This distribution can also be described using gamma distribution function \citep[e.g.][]{handbook} which starts from magnitude $m\approx16$, peaks at $m\approx14$ and spans upto the saturation limit of $m\approx6$. Prominence star clusters that emerge from crowding field stars implies that member stars follow the same distribution, but with brighter mean. The result of this scheme is presented in \autoref{fig:monte}.

Proper motions distribution of the field stars is basically governed by intrinsic motions of different stellar populations (e.g. thin and thick disk stars), differential rotation of the Galaxy and also solar motion. Thus, the distribution depend on the direction of the observational field. On the other hand, the distribution of member stars depends on the tangential motion of the cluster itself. Since the intrinsic dispersion of stars inside open cluster is relatively small $\sim1$ km/s and observed from parsecs distance, this dispersion is overwhelmed by measurement errors \citep{nunez07}. Departing from these assumptions, we generate proper motions data of non-members using random number with normal distribution, centered at $(\overline{\mu}_{x,f},\overline{\mu}_{y,f})=(0,0)$ with dispersion of $\sigma=8$ mas/year. This value is comparable to the parametric solutions of $\sigma_f$ obtained by \cite{krone2010} from the analysis of nine clusters with PM2000 proper motions data \citep{ducourant06}. On the other hand, the proper motions of all member stars are identical, equal to cluster's proper motions.

To obtain more realistic model, measurement errors are assigned  for every proper motions data. The magnitude of proper motions errors grow exponentially as the magnitude increase (e.g. dimmer stars have larger error). The upper limit of this errors is 10 mas/year while the minimum value is specified by input parameter $a_{\epsilon}=[0,6]$ which will be explained later. Lower-right plot of \autoref{fig:monte} shows error profiles with different error parameters.

\subsection{Free Parameters}
During the simulations, there are three varied parameters governing 1000 different synthetic clusters to be analyzed. The first parameter is the number ratio of cluster members and field stars ($N_c/N_f$) which ranges from 0.1 to 2.0, equivalently $N_c\approx[60,1200]$ if the number of field stars is set to be constant $N_f=600$. This parameter determines how scarce the cluster members are. The second parameter defines centroid position of member stars' kinematic. \citet{cano90} stated that the distance between two populations' centroid influences the accuracy of proper motions determination. The third parameter is the proper motions errors that strongly depend on the apparent magnitude of the stars. In this simulation, the errors grow exponentially as the magnitude increase. \autoref{eq:error} defines proper motions errors as the function of magnitude. 

\begin{equation}
\epsilon_{\mu}=a_{\epsilon}+(10-a_{\epsilon})\exp\left(\frac{m}{2}-8\right),
\label{eq:error}
\end{equation}
with $a_{\epsilon}=[0,6]$ characterizes the whole function. This function is inspired by the exponential profile of measurement error from UCAC4 data which come from several catalogues with slightly different accuracies \citep{zacharias13}. To simulate the measurement error, every proper motion data is replaced by a random number taken from Gaussian distribution with the mean equal to the original proper motion and the standard deviation equal to the proper motion error.

\section{Results and Analysis}
\label{sec:result}
For each simulated model, we perform standard parametric method as described by \citet{zhao90} and non-parametric (both KDE and \bkdee) kinematic analysis to determine both membership probability of any star and cluster's proper motion. The output of this process which are membership probability and distribution center will be analysed as follow.

\subsection{Membership Probability}
\begin{figure*}[ht]
\centering
\includegraphics[width=0.75\textwidth]{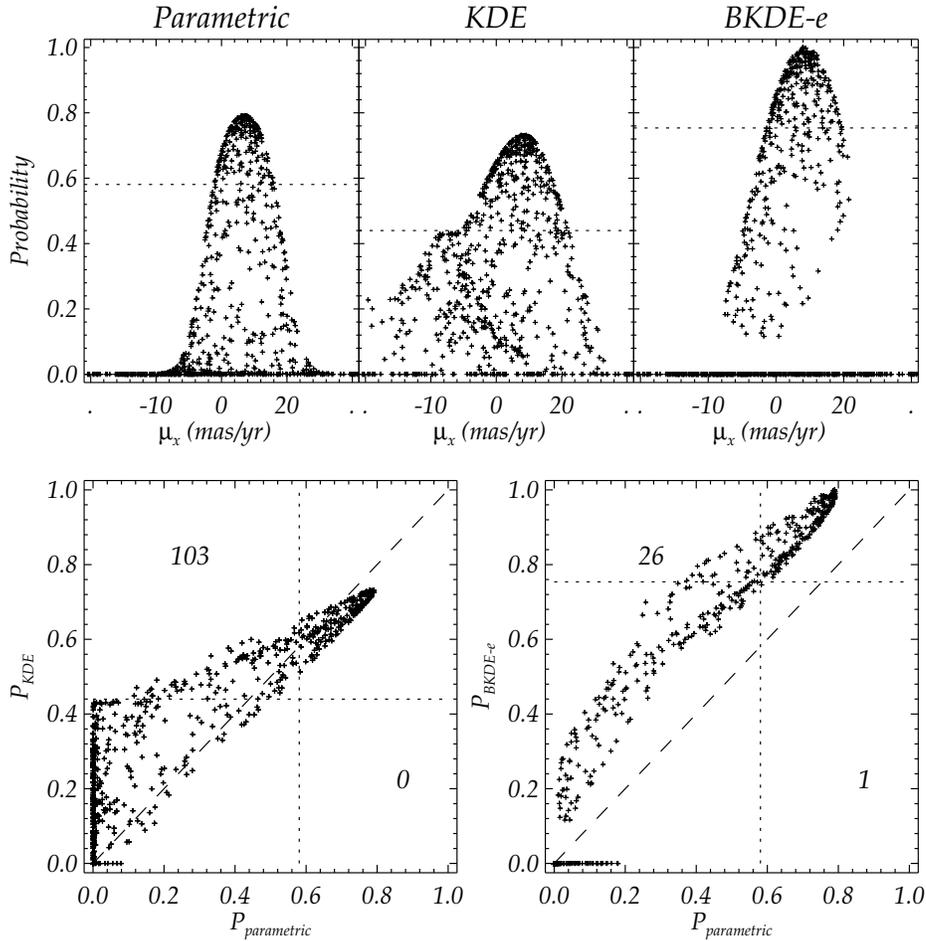}
\caption{Three upper figures show membership probability of the simulated stars as function of proper motions in $x$ direction, derived by three mentioned methods. Non-parametric probability may show more than one peaks (more obvious in KDE analysis). As consequence, non-parametric probability for several stars are higher than their parametric probability (lower figure). Dotted line in those figure mark $2\sigma$ limit of determined probability.}
\label{fig:monte-prob}
\end{figure*}

\newcommand{\flow}{\hat{f}_{\text{lim}}}

As mentioned by previous authors \citep{cano90,galadi98,nunez07}, non-parametric membership probability of any star can be calculated using the following equation:
\begin{equation}
P_i=\left\{
\begin{array}{lr}
\dfrac{\hat{f}_{c+f,i}-\hat{f}_{f,i}}{\hat{f}_{c+f,i}}, & \text{if } \hat{f}_{c+f,i}>\flow\\
0, & \text{if } \hat{f}_{c+f,i}\leq\flow
\end{array}\right.
\label{proba}
\end{equation}
where $\hat{f}_{c+f}$ and $\hat{f}_c$ define the density estimation of infield and out-field stars, while $\flow$ sets the bottom limit to exclude insignificant density estimate at the tail of distribution. In this study, $\flow\approx10^{-4}$ that corresponds to $2.5\sigma$ normal distribution is used. This exclusion is necessary to avoid overestimation of membership probability due to density fluctuation at the edge of distribution, e.g. $P\approx1$ when $\hat{f}_f\approx0$.

Overestimation of non-parametric probability (as shown in \autoref{fig:monte-prob}) can be treated through at least two ways. Firstly by increasing the calculation lower limit ($\flow$) that automatically yields zero probability at the edge of distribution. Second treatment is to use more out-field stars in order to statistically improve the out-field distribution profile ($\hat{f}_{f}$). This procedure comes as alternative treatment as very low $\hat{f}_{f}$ can be driven by under-sampled data. The implementation of those treatment have not been investigated deeply in this study since the overestimation does not really influence the proper motions determination of any cluster.

\subsection{Confidence Level}
The achieved membership probability, both parametric and non parametric, can be categorized into several confidence level according to the cumulative density function:
\begin{equation}
CDF_j=\sum_{i=1}^jP_{\text{sort},i},
\end{equation}
where $P_{\text{sort},i}$ represents membership probability, sorted increasingly. By assuming Gaussian probability function, $1\sigma$ members have $CDF\geq31.73\%$, $2\sigma$ have $CDF\geq4.55\%$, while $3\sigma$ have $CDF\geq0.27\%$\footnote{\emph{Two-tails Gaussian test}.}.

From this category, list of cluster members can be composed, for example by assigning $2\sigma$ members as the probable cluster members. As shown in the lower panels of \autoref{fig:monte-prob}, non-parametric methods (both KDE and {\bkdee}) often assign higher probability compared to parametric results. The difference is that in KDE, the data with low parametric probability tend to deviate more. In {\bkdee} analysis, there are more $2\sigma$ members but false assignment of less probable members are suppressed. {\bkdee} assigns probability limit more loosely compared to the parametric method as it accounts asymmetric distribution function or even the existence of multi-modality or blended distributions.

\begin{figure*}[ht]
\centering
\includegraphics[width=0.9\textwidth]{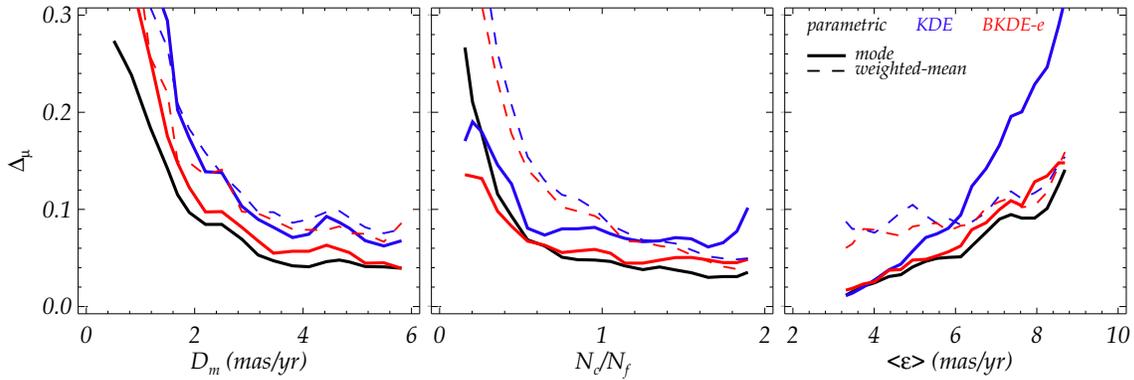}
\caption{Dimensionless accuracy parameter of proper motions determination ($\Delta_{\mu}$) as function of three free parameters: centroid separation (left), $N_c/N_f$ ratio (middle) and average measurement errors (right). The lines represent median value of accuracy parameter for each abscissa bin (analogous to moving average). Red, blue, and black color correspond to {\bkdee}, KDE and parametric analysis respectively. Solid and dashed line correspond to mode and weighted mean procedures for non-parametric analysis.}
\label{fig:dp}
\end{figure*}

\subsection{Average Proper Motion}
Beside the assignment of membership probability, kinematic analysis will give the centroid of proper motions distribution of cluster stars and field stars. The centroid of cluster stars corresponds to the average proper motions or the bulk-motion of the cluster.

The location of the centroid, both in $\mu_x$ and $\mu_y$ direction, are the main parameters of parametric equations \citep{zhao90,krone2010}. In such way, the cluster's proper motions is achieved as the algorithm reaches the convergence solutions. On the other hand, non-parametric method will yield density estimation $\phi_c=\phi_{c+f}-\phi_f$ where the mode location of $\phi_c$ can be considered as the average proper motion. But, determination of the mode it self is a tricky process. \citet{nunez07} used bin location with the highest $\phi_c$ as the indicator of cluster's bulk-motion with the size of the bin as its uncertainty.

In this study, the following procedures are employed:
\begin{itemize}
\item Based on the established $\hat{f}_c$, data points with $\hat{f}_c\geq90\%\times\hat{f}_{c,\text{max}}$ are assigned as the elite data used for the next steps. The probability limit of 90\% is chosen based on the assumption that the attained elite data represent the mode of the distribution.
\item The mode or the average proper motions is the average value of elite data points:
\begin{equation}
\overline{\mu}=\dfrac{1}{n_{\text{elite}}}\sum_{i=1}^{n_{\text{elite}}}\mu_{\text{elite},i}.
\end{equation}
\item The standard deviation of elite data points is considered as the proper motion's uncertainty:
\begin{equation}
\sigma_{\overline{\mu}}^2=\dfrac{1}{n_{\text{elite}}-1}\sum_{i=1}^{n_{\text{elite}}}(\mu_{\text{elite},i}-\overline{\mu})^2.
\end{equation}
\end{itemize}

\begin{table}[ht]
\small
\centering
\caption{Statistics of cluster proper motions determination accuracy parameter ($\Delta_{\mu}$) for each method summarized by its mean and median.}
\label{tab:median}
\vskip5pt
{\renewcommand{\arraystretch}{1.2}
\begin{tabular}{|l|cc|cc|}
\hline
\multirow{2}{1in}{\textbf{Method}} & \multicolumn{2}{|c|}{$\Delta_{\mu}$}  & \multicolumn{2}{|c|}{Accuracy} \\
& \textbf{Mean} & \textbf{Med} & \textbf{Mean} & \textbf{Med} \\
\hline
Parametric & 0.120 & 0.056 & 88\% & 94\%\\
Mode: KDE & 0.204 & 0.093 & 80\% & 91\% \\
Mode: \bkdee & 0.135 & 0.068 & 86\% & 93\% \\
w-mean: KDE & 0.209 & 0.120 & 79\% & 88\% \\
w-mean: \bkdee & 0.190 & 0.095 & 81\% & 91\% \\
\hline
\end{tabular}}
\end{table}

The usage of mode as indicator of cluster's bulk motion in non-parametric methods have positive and negative features. In this procedure, skewness or the wide wing of distribution will not affect the final result. But, subtraction in \autoref{proba} may slightly deform the shape of $\hat{f}_c$ and shift the location of the mode.

Another alternative procedure to determine the proper motions is using probability-weighted average:
\begin{equation}
\overline{\mu}=\dfrac{\sum_{i=1}^{n}\mu_iP_i}{\sum_{i=1}^{n} P_i},
\end{equation}
with membership probability $P_i$. This procedure is very sensitive to the membership probability determination. Then, only $2\sigma$ members are included in calculation.

\begin{table*}[ht]
\centering
\caption{Proper motion of NGC 2682 obtained by several authors. Among these authors, \cite{bellini2010} used CFHT and LBT observations, \cite{krone2010} used PM2000 \citep{ducourant06}, \cite{nunez07} used UCAC2 \citep{zacharias04}, while the rest directly used \emph{Tycho-2} catalog \citep{hog00}.}
\label{compare}
\vskip5pt
{\renewcommand{\arraystretch}{1.2}
\begin{tabular}{|c|c|l|l|}
\hline
$\mu_{\alpha}\cos{\delta}$ & $\mu_{\delta}$ & Referece & Data Source\\
\hline
$-9.94\pm0.85$ & $-4.92\pm0.88$ & this study & UCAC4\\
$-9.6\pm1.1$ & $-3.7\pm0.8$ & \cite{bellini2010} & CFHT \& LBT obs.\\
$-8.32\pm0.07$ & $-5.65\pm0.07$ & \cite{krone2010} & PM2000\\
$-7.87\pm0.61$ & $-5.60\pm0.59$ & \cite{frinchaboy08} & Tycho-2\\
$-7.1\pm0.8$ & $-7.6\pm0.4$ & \cite{nunez07} & UCAC2\\
$-8.31\pm0.26$ & $-4.81\pm0.22$ & \cite{kharchenko05} & Tycho-2\\
$-8.62\pm0.28$ & $-6.00\pm0.28$ & \cite{dias02p} & Tycho-2\\
\hline
\end{tabular}}
\end{table*}

\subsection{Accuracy of the proper motions Determination}
To evaluate and compare the accuracy of three methods, an accuracy parameter $\Delta_{\mu}$ is defined as follow:
\begin{equation}
\Delta_{\mu}=\dfrac{|\overline{\mu}-\overline{\mu}_0|}{D_{\mu}},
\end{equation}
where $\overline{\mu}$ is mean proper motions from parametric and non-parametric analysis, using mode and weighted mean procedures, while $\overline{\mu}_0$ represents true value of the proper motions and $D_{\mu}$ represents the distance of cluster and field stars distribution in VPD (centroid separation). This parameter is equal to 0 for accurate determination or $\gtrsim1$ for inaccurate determination. Plots of the $\Delta_{\mu}$ from 1000 simulated models are shown in \autoref{fig:dp}. 

\begin{description}
\item[$\Delta_{\mu}$ versus $D_{\mu}$]:
As expected, the accuracy of proper motions determinations increase as the centroid of two populations become more separated. The relative error, $\Delta_{\mu}$ decrease as $D_{\mu}$ increase. It is shown that the weighted mean solutions of non-parametric analysis are not better than mode solution of {\bkdee}. This latter solution is as good as parametric solution.

\item[$\Delta_{\mu}$ versus $N_c/N_f$]:
Variation of $\Delta_{\mu}$ as function of $N_c/N_f$ achieved from the simulation is in line with \citet{cano90} that the accuracy increase as the ratio increase. Cluster's centroid can be indentified more easily as the cluster becomes more populous. Similar behaviour of mode solution of {\bkdee}, which is superior to KDE and weighted mean solutions, is also observed. Lower accuracy of weighted mean solutions come from overestimated membership probability as discussed before.

\item[$\Delta_{\mu}$ versus $\langle{\epsilon}\rangle$]:
The accuracy of proper motions determination declines monotonically as measurement errors grows. It is trivial, but the inferiority of mode solution of KDE becomes more clearly as shown in \autoref{fig:dp}. At $\langle{\epsilon}\rangle\sim7$ the relative error of KDE mode solution is almost twice of another solutions. Nevertheless, simulation does not show that {\bkdee} serves better result compared to parametric approach, even in large measurement errors.
\end{description}

Median and mean value from  $\Delta_{\mu}$ for each method are also enlisted in \autoref{tab:median}. Based on those distributions, it can be summarized that parametric method has the highest accuracy (on average $\sim88\%$). The second highest accuracy comes from mode solution in {\bkdee}, with average accuracy $\sim86\%$. KDE mode solution has accuracy of $\sim80\%$, almost similar with weighted mean solution. Same rank happens to median $\Delta_{\mu}$. This result shows that {\bkdee} method is eligible to be used in the future works.

\begin{figure*}[ht]
\centering
\includegraphics[width=0.65\textwidth]{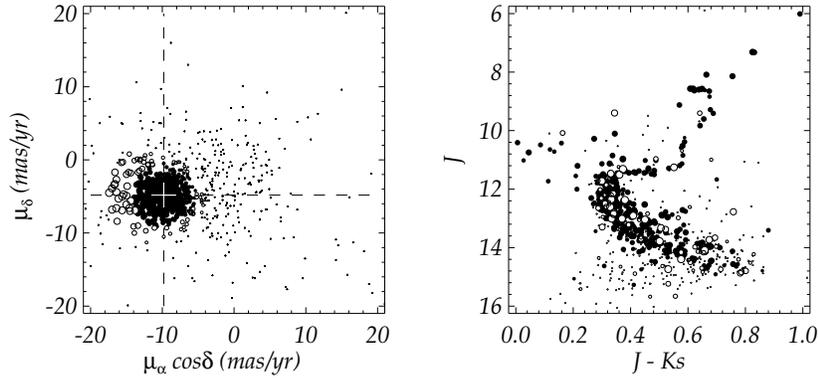}
\caption{Vector point (left) and color-magnitude (right) of star cluster NGC 2682 (M 67). Membership probability of each star is denoted by circle with various size, proportional to the probability. Dashed lines in VPD mark the proper motion of the cluster obtained using \bkdee.}
\label{ngc2682}
\end{figure*}

\section{Application to NGC 2682}
It is of interest to apply {\bkdee} to the real cluster data. As the early comparison, we choose NGC 2682 (M 67) as the subject cluster. This cluster can be a candidate as the birth place of the Sun according to the similarities in metallicity and the predicted position at 4.5 Gyr ago though controversy still exist (see \citet{brown2010} and also \citet{pichardo2012}). Accurate determination of the actual cluster's motion in space provides an input parameter to trace back the cluster's position in the past. This will lead to the conclusion whether the historical position of the Sun coincided with NGC 2682.

To determine the proper motion of NGC 2682, we use proper motion from the fourth US Naval Observatory CCD Astrograph Catalog (UCAC4) that includes more than 105 million stars with measured proper motion up to $R$-band magnitude of $\sim16$ \citep{zacharias13}. Typical proper motion error in proper motion ranges from 1 to 10 mas/yr, but we only use stars with error less than 7 mas/yr for further analysis. Stars within tidal radius of $0.31'$ \citep{kharchenko05} around the cluster's center are considered as \emph{in-field} stars while the stars within ring area $0.35'<r<0.43'$ are \emph{out-field} stars. For the case of concentrated-cluster NGC 2682, the inappropriate choice of in-field and out-field radii does not significantly affect the obtained cluster proper motion, but overestimation of the radii could increase the rate of false membership assignment as alerted by \cite{sanchez09}. On the other hand, underestimation of out-field radius misleads to a deviated density estimate of field stars kinematics. This problem is not significant for rich or concentrated clusters but not for diffuse clusters which could be overcame by employing a slightly larger in-field and out-field radii. In this way, false member assignments need to be pruned by photometric analysis, e.g. through CMD.

\autoref{ngc2682} depicts membership analysis using {\bkdee}. In this case, cluster's members are relatively easy to be separated from the field stars. Based on the mode of cluster's density estimate ($\hat{f}_c$), we obtain $\mu_{\alpha}\cos{\delta}=-9.94\pm0.85$ mas/yr and $\mu_{\delta}=-4.92\pm0.88$ mas/yr. These results are comparable to the recent studies as summarized in \autoref{compare}. However, non-systematic deviation of the obtained proper motions, less than 0.1 mas/yr, is observed when the in-field and out-field radii are varied between 0.7 to 1.3 times tidal radius.

The result of decontamination process is shown in color-magnitude diagram where giant branch and blue-straggler populations are clearly identified as cluster members. However, an asymmetric non-parametric membership probability is observed in vector point diagram where stars located away from field stars distribution centre are considered as members with systematically higher membership probability.

Beyond the case of NGC 2682, UCAC4 provides a large number of proper motions for cluster kinematic analysis. This is the subject of study in the near future.

\section{Conclusions}
\label{sec:conclusions}

In the present study, we employ non-parametric \emph{Binned Kernel Density Estimator} for open cluster membership analysis using proper motions data. Some modifications are implemented to this method in order to account the effect of measurement errors through numerical approach (\bkdee). A thousand of Monte Carlo models are simulated to evaluate the behaviour of {\bkdee} compared to the basic KDE and parametric approach. As the result, proper motions determination from {\bkdee} analysis using mode searching algorithm is better than the KDE solutions. {\bkdee} solutions are as accurate as parametric solutions. Simulation also shows that the accuracy of proper motions determination decreases as the member stars far outnumbered by field stars, though steeper decline is observed in accuracy versus centroid distance plot. These conclusions quantitatively support the previous statements of \cite{cano90}.

By including measurement errors in the calculation, the mode location from the resulting density estimate is less sensitive to non-physical or stochastic fluctuation as compared to ordinary KDE that excludes measurement errors. For the typical mean measurement error of 7 mas/yr, {\bkdee} suppresses the potential of miscalculation by a factor of two compared to KDE. With median accuracy of about 93\%, {\bkdee} based method has comparable accuracy with respect to parametric method.

Additionally, {\bkdee} is also implemented to the real proper motions data gathered from UCAC4 around the field of NGC 2682. Non-parametric analysis yields cluster's proper motions of  $\mu_{\alpha}\cos\delta=-9.94\pm0.85$ mas/yr and $\mu_{\delta}=-4.92\pm0.88$ mas/yr which are comparable to other studies.

Although {\bkdee} method does not overtake parametric approach, it provides flexibility in terms that it can be employed to astrometric and photometric data \citep{cano90,galadi98}. Stars distribution of diffuse open cluster in space can be treated with non-parametric approach up to a certain limit of density contrast. Similarly, photometric data in Color Magnitude Diagram can also be decontaminated through non-parametric process. The resulting photometric density estimate of this analysis is also expected to form a fiducial sequence for further physical analysis such as isochrone fitting.

Moreover, non-parametric approach is expected as the efficient method for overlapping clusters or even binary star clusters \citep{marcos09,priyatikanto13}. Detected multi-modality over Vector Point Diagram may confirms the exisence of substructure or even the physical interaction between two clusters.

\section*{Acknowledgements}
Authors are grateful to the anonymous referee for his/her constructive comments. R. Priyatikanto also indebted to Tim Pembina Olimpiade Astronomi Indonesia (TPOA) for its continuous support.

The final publication is available at Springer via \url{http://dx.doi.org/10.1007/s10509-014-2137-y}
\bibliographystyle{spr-mp-nameyear-cnd}
\bibliography{priyatikanto.bib}
\end{document}

%% file: priyatikanto-fig1a.tex
\begin{tikzpicture}[scale=1,>=latex]
	\draw (-1,3) node {(a)};
	\fill [red!60] (3,0) rectangle (4,2.5);
	\fill [green!60] (4,0) rectangle (6,2.5);
	\fill (3.5,2) node {$w_{k,c}$};
	\fill (5,0.5) node {$w_{k,b}$};
	\draw [->,thick] (4.5,0.5) to [out=-180, in=30] (3.2,0.1);
	\draw [->,thick] (4,2.0) to [out=0, in=105] (5.9,0.2);

	\draw (0,0) -- (0,3);
	\draw (3,0) -- (3,3);
	\draw (6,0) -- (6,3);
	\draw (9,0) -- (9,3);
	\draw [<->] (3,2.8) -- (6,2.8) node [pos=0.5,above] {$H$};

	\draw [line width=2pt] (0,0) -- (10,0);
	\draw [dashed] (4,0) -- (4,2.5);
	\fill (4,0) circle (3pt) node[below] {$x$};
	\fill (0,0) circle (3pt) node[below] {$x_{k,a}$};
	\fill (3,0) circle (3pt) node[below] {$x_{k,b}$};
	\fill (6,0) circle (3pt) node[below] {$x_{k,c}$};
	\fill (9,0) circle (3pt) node[below] {$x_{k,d}$};
\end{tikzpicture}

%% file: priyatikanto-fig1b.tex
\begin{tikzpicture}[scale=1,>=latex]
	\draw (-1,3) node {(b)};
	\begin{scope}
		\clip (1,0) to [out=10, in=-115] (3,1.4) to [out=65, in=180] (4,2.5) 
			to [out=0, in=115] (5,1.4) to [out=-65, in=170] (7,0) -- cycle;
		\clip (0,0) rectangle (3,4);
		\fill[color=blue!60] (0,0) rectangle (10,4);
	\end{scope}
	
	\begin{scope}
		\clip (1,0) to [out=10, in=-115] (3,1.4) to [out=65, in=180] (4,2.5) 
			to [out=0, in=115] (5,1.4) to [out=-65, in=170] (7,0) -- cycle;
		\clip (3,0) rectangle (6,4);
		\fill[color=red!60] (0,0) rectangle (10,4);
	\end{scope}

	\begin{scope}
		\clip (1,0) to [out=10, in=-115] (3,1.4) to [out=65, in=180] (4,2.5) 
			to [out=0, in=115] (5,1.4) to [out=-65, in=170] (7,0) -- cycle;
		\clip (6,0) rectangle (9,4);
		\fill[color=green!60] (0,0) rectangle (10,4);
	\end{scope}

	\draw (0,0) -- (0,3);
	\draw (3,0) -- (3,3);
	\draw (6,0) -- (6,3);
	\draw (9,0) -- (9,3);
	\draw [dashed,->] (1,3) -- (1,2.5);
	\draw [dashed,->] (7,3) -- (7,2.5);
	\draw (1,3) node [above] {$x-3\sigma$};
	\draw (7,3) node [above] {$x+3\sigma$};

	\draw [line width=2pt] (0,0) -- (10,0);
	\draw [dashed] (4,0) -- (4,3) node[above] {$x$};
	\fill (0,0) circle (3pt) node[below] {$x_{k,a}$};
	\fill (3,0) circle (3pt) node[below] {$x_{k,b}$};
	\fill (6,0) circle (3pt) node[below] {$x_{k,c}$};
	\fill (9,0) circle (3pt) node[below] {$x_{k,d}$};
\end{tikzpicture}

%% file: priyatikanto-ms.bbl
\begin{thebibliography}{30}
\ifx \bisbn   \undefined \def \bisbn  #1{ISBN #1}\fi
\ifx \binits  \undefined \def \binits#1{#1} \fi
\ifx \bauthor  \undefined \def \bauthor#1{#1} \fi
\ifx \batitle  \undefined \def \batitle#1{#1} \fi
\ifx \bjtitle  \undefined \def \bjtitle#1{#1}\fi
\ifx \bvolume  \undefined \def \bvolume#1{\textbf{#1}}\fi
\ifx \byear  \undefined \def \byear#1{#1} \fi
\ifx \bissue  \undefined \def \bissue#1{#1} \fi
\ifx \bfpage  \undefined \def \bfpage#1{#1} \fi
\ifx \blpage  \undefined \def \blpage #1{#1} \fi
\ifx \burl  \undefined \def \burl#1{\textsf{#1}} \fi
\ifx \doiurl  \undefined \def \doiurl#1{\textsf{#1}} \fi
\ifx \betal  \undefined \def \betal{\textit{et al.}} \fi
\ifx \binstitute  \undefined \def \binstitute#1{#1} \fi
\ifx \binstitutionaled  \undefined \def \binstitutionaled#1{#1} \fi
\ifx \bctitle  \undefined \def \bctitle#1{#1} \fi
\ifx \beditor  \undefined \def \beditor#1{#1} \fi
\ifx \bpublisher  \undefined \def \bpublisher#1{#1} \fi
\ifx \bbtitle  \undefined \def \bbtitle#1{#1} \fi
\ifx \bedition  \undefined \def \bedition#1{#1} \fi
\ifx \bseriesno  \undefined \def \bseriesno#1{#1} \fi
\ifx \blocation  \undefined \def \blocation#1{#1} \fi
\ifx \bsertitle  \undefined \def \bsertitle#1{#1} \fi
\ifx \bsnm \undefined \def \bsnm#1{#1} \fi
\ifx \bsuffix \undefined \def \bsuffix#1{#1} \fi
\ifx \bparticle \undefined \def \bparticle#1{#1} \fi
\ifx \barticle \undefined \def \barticle#1{#1} \fi
\ifx \bconfdate \undefined \def \bconfdate #1{#1} \fi
\ifx \botherref \undefined \def \botherref #1{#1} \fi
\ifx \url \undefined \def \url#1{\textsf{#1}} \fi
\ifx \bchapter \undefined \def \bchapter#1{#1} \fi
\ifx \bbook \undefined \def \bbook#1{#1} \fi
\ifx \bcomment \undefined \def \bcomment#1{#1} \fi
\ifx \oauthor \undefined \def \oauthor#1{#1} \fi
\ifx \citeauthoryear \undefined \def \citeauthoryear#1{#1} \fi
\ifx \endbibitem  \undefined \def \endbibitem {}\fi
\ifx \bconflocation  \undefined \def \bconflocation#1{#1} \fi
\ifx \arxivurl  \undefined \def \arxivurl#1{\textsf{#1}} \fi

\bibitem[\protect\citeauthoryear{{Balaguer-N\'un\~ez} et~al.}{2007}]{nunez07}
\begin{barticle}
\bauthor{\bsnm{{Balaguer-N\'un\~ez}}, \binits{L.}},
\bauthor{\bsnm{{Galad\'i-Enr\'iquez}}, \binits{D.}},
\bauthor{\bsnm{Jordi}, \binits{C.}}:
\bjtitle{\aap}
\bvolume{470},
\bfpage{585}
(\byear{2007})
\end{barticle}
\endbibitem

\bibitem[\protect\citeauthoryear{Bellini et~al.}{2010}]{bellini2010}
\begin{barticle}
\bauthor{\bsnm{Bellini}, \binits{A.}},
\bauthor{\bsnm{Bedin}, \binits{L.R.}},
\bauthor{\bsnm{Pichardo}, \binits{B.}},
\bauthor{\bsnm{Moreno}, \binits{E.}},
\bauthor{\bsnm{Allen}, \binits{C.}},
\bauthor{\bsnm{Piotto}, \binits{G.}},
\bauthor{\bsnm{Anderson}, \binits{J.}}:
\bjtitle{\aap}
\bvolume{513},
\bfpage{51}
(\byear{2010})
\end{barticle}
\endbibitem

\bibitem[\protect\citeauthoryear{Brown et~al.}{2010}]{brown2010}
\begin{barticle}
\bauthor{\bsnm{Brown}, \binits{A.G.A.}},
\bauthor{\bsnm{Portegies-Zwart}, \binits{S.F.}},
\bauthor{\bsnm{Bean}, \binits{J.}}:
\bjtitle{\mnras}
\bvolume{407},
\bfpage{458}
(\byear{2010})
\end{barticle}
\endbibitem

\bibitem[\protect\citeauthoryear{{Cabrera-Ca\~no} and Alfaro}{1985}]{cano85}
\begin{barticle}
\bauthor{\bsnm{{Cabrera-Ca\~no}}, \binits{J.}},
\bauthor{\bsnm{Alfaro}, \binits{E.J.}}:
\bjtitle{\aap}
\bvolume{150},
\bfpage{298}
(\byear{1985})
\end{barticle}
\endbibitem

\bibitem[\protect\citeauthoryear{{Cabrera-Ca\~no} and Alfaro}{1990}]{cano90}
\begin{barticle}
\bauthor{\bsnm{{Cabrera-Ca\~no}}, \binits{J.}},
\bauthor{\bsnm{Alfaro}, \binits{E.J.}}:
\bjtitle{\aap}
\bvolume{235},
\bfpage{94}
(\byear{1990})
\end{barticle}
\endbibitem

\bibitem[\protect\citeauthoryear{{de la Fuente Marcos} and {de la Fuente
  Marcos}}{2009}]{marcos09}
\begin{barticle}
\bauthor{\bsnm{{de la Fuente Marcos}}, \binits{R.}},
\bauthor{\bsnm{{de la Fuente Marcos}}, \binits{C.}}:
\bjtitle{\aap}
\bvolume{500},
\bfpage{13}
(\byear{2009})
\end{barticle}
\endbibitem

\bibitem[\protect\citeauthoryear{Dias et~al.}{2002}]{dias02}
\begin{barticle}
\bauthor{\bsnm{Dias}, \binits{W.S.}},
\bauthor{\bsnm{Alessi}, \binits{B.S.}},
\bauthor{\bsnm{Moitinho}, \binits{A.}},
\bauthor{\bsnm{L\'{e}pine}, \binits{J.R.D.}}:
\bjtitle{\aap}
\bvolume{389},
\bfpage{871}
(\byear{2002})
\end{barticle}
\endbibitem

\bibitem[\protect\citeauthoryear{Dias et~al.}{2006}]{dias06}
\begin{barticle}
\bauthor{\bsnm{Dias}, \binits{W.S.}},
\bauthor{\bsnm{Assafin}, \binits{M.}},
\bauthor{\bsnm{Fl\'orio}, \binits{V.}},
\bauthor{\bsnm{Alessi}, \binits{B.S.}},
\bauthor{\bsnm{L\'ibero}, \binits{V.}}:
\bjtitle{\aap}
\bvolume{446},
\bfpage{949}
(\byear{2006})
\end{barticle}
\endbibitem

\bibitem[\protect\citeauthoryear{Dias et~al.}{2002}]{dias02p}
\begin{barticle}
\bauthor{\bsnm{Dias}, \binits{W.S.}},
\bauthor{\bsnm{L\'epine}, \binits{J.R.D.}},
\bauthor{\bsnm{Alessi}, \binits{B.S.}}:
\bjtitle{\aap}
\bvolume{388},
\bfpage{168}
(\byear{2002})
\end{barticle}
\endbibitem

\bibitem[\protect\citeauthoryear{Ducourant et~al.}{2006}]{ducourant06}
\begin{barticle}
\bauthor{\bsnm{Ducourant}, \binits{C.}},
\bauthor{\bsnm{Le~Campion}, \binits{J.F.}},
\bauthor{\bsnm{Rapaport}, \binits{M.}}, \betal:
\bjtitle{\aap}
\bvolume{448},
\bfpage{1235}
(\byear{2006})
\end{barticle}
\endbibitem

\bibitem[\protect\citeauthoryear{Frinchaboy and Majewski}{2008}]{frinchaboy08}
\begin{barticle}
\bauthor{\bsnm{Frinchaboy}, \binits{P.M.}},
\bauthor{\bsnm{Majewski}, \binits{S.R.}}:
\bjtitle{\aj}
\bvolume{136},
\bfpage{118}
(\byear{2008})
\end{barticle}
\endbibitem

\bibitem[\protect\citeauthoryear{Galadi-Enriquez et~al.}{1998}]{galadi98}
\begin{barticle}
\bauthor{\bsnm{Galadi-Enriquez}, \binits{D.}},
\bauthor{\bsnm{Jordi}, \binits{C.}},
\bauthor{\bsnm{Trullols}, \binits{E.}}:
\bjtitle{\aap}
\bvolume{337},
\bfpage{125}
(\byear{1998})
\end{barticle}
\endbibitem

\bibitem[\protect\citeauthoryear{H{\o}g et~al.}{2000}]{hog00}
\begin{barticle}
\bauthor{\bsnm{H{\o}g}, \binits{E.}},
\bauthor{\bsnm{Fabricius}, \binits{C.}},
\bauthor{\bsnm{Makarov}, \binits{V.}},
\bauthor{\bsnm{Urban}, \binits{S.}},
\bauthor{\bsnm{Corbin}, \binits{T.}},
\bauthor{\bsnm{Wycoff}, \binits{G.}},
\bauthor{\bsnm{Bastian}, \binits{U.}},
\bauthor{\bsnm{Schwedendiek}, \binits{P.}},
\bauthor{\bsnm{Wicenec}, \binits{A.}}:
\bjtitle{\aap}
\bvolume{355},
\bfpage{27}
(\byear{2000})
\end{barticle}
\endbibitem

\bibitem[\protect\citeauthoryear{Kharchenko et~al.}{2005}]{kharchenko05}
\begin{barticle}
\bauthor{\bsnm{Kharchenko}, \binits{N.V.}},
\bauthor{\bsnm{Piskunov}, \binits{A.E.}},
\bauthor{\bsnm{R\"{o}ser}, \binits{S.}},
\bauthor{\bsnm{Schilbach}},
\bauthor{\bsnm{E.}},
\bauthor{\bsnm{Scholz}, \binits{R.-D.}}:
\bjtitle{\aap}
\bvolume{483},
\bfpage{1163}
(\byear{2005})
\end{barticle}
\endbibitem

\bibitem[\protect\citeauthoryear{Krishnamoorthy}{2006}]{handbook}
\begin{bbook}
\bauthor{\bsnm{Krishnamoorthy}, \binits{K.}}:
\bbtitle{Handbook of Statistical Distributions with Applications}.
\bpublisher{{Chapman} \& {Hall}/{CRC}},
\blocation{London}
(\byear{2006})
\end{bbook}
\endbibitem

\bibitem[\protect\citeauthoryear{{Krone-Martins} et~al.}{2010}]{krone2010}
\begin{barticle}
\bauthor{\bsnm{{Krone-Martins}}, \binits{A.}},
\bauthor{\bsnm{Soubiran}, \binits{C.}},
\bauthor{\bsnm{Ducourant}, \binits{C.}},
\bauthor{\bsnm{Teixeira}, \binits{R.}},
\bauthor{\bsnm{{Le Campion}}, \binits{J.F.}}:
\bjtitle{\aap}
\bvolume{516},
\bfpage{3}
(\byear{2010})
\end{barticle}
\endbibitem

\bibitem[\protect\citeauthoryear{L\'{e}pine et~al.}{2011}]{lepine11}
\begin{barticle}
\bauthor{\bsnm{L\'{e}pine}, \binits{J.R.D.}},
\bauthor{\bsnm{Cruz}, \binits{P.}},
\bauthor{\bsnm{{Scarano Jr}}, \binits{S.}},
\bauthor{\bsnm{Barros}, \binits{D.A.}},
\bauthor{\bsnm{Dias}, \binits{W.S.}},
\bauthor{\bsnm{Pomp\'{e}ia}, \binits{L.}},
\bauthor{\bsnm{Andrievsky}, \binits{S.M.}},
\bauthor{\bsnm{Carraro}, \binits{G.}},
\bauthor{\bsnm{Famaey}, \binits{B.}}:
\bjtitle{\mnras}
\bvolume{417},
\bfpage{698}
(\byear{2011})
\end{barticle}
\endbibitem

\bibitem[\protect\citeauthoryear{Pichardo et~al.}{2012}]{pichardo2012}
\begin{barticle}
\bauthor{\bsnm{Pichardo}, \binits{B.}},
\bauthor{\bsnm{Moreno}, \binits{E.}},
\bauthor{\bsnm{Allen}, \binits{C.}},
\bauthor{\bsnm{Bedin}, \binits{L.R.}},
\bauthor{\bsnm{Bellini}, \binits{A.}},
\bauthor{\bsnm{Pasquini}, \binits{L.}}:
\bjtitle{\aj}
\bvolume{143},
\bfpage{73}
(\byear{2012})
\end{barticle}
\endbibitem

\bibitem[\protect\citeauthoryear{Piskunov et~al.}{2006}]{piskunov06}
\begin{barticle}
\bauthor{\bsnm{Piskunov}, \binits{A.E.}},
\bauthor{\bsnm{Kharchenko}, \binits{N.V.}},
\bauthor{\bsnm{R\"{o}ser}, \binits{S.}},
\bauthor{\bsnm{Schilbach}, \binits{E.}},
\bauthor{\bsnm{Scholz}, \binits{R.-D.}}:
\bjtitle{\aap}
\bvolume{445},
\bfpage{545}
(\byear{2006})
\end{barticle}
\endbibitem

\bibitem[\protect\citeauthoryear{Priyatikanto et~al.}{2014}]{priyatikanto13}
\begin{barticle}
\bauthor{\bsnm{Priyatikanto}, \binits{R.}},
\bauthor{\bsnm{Arifyanto}, \binits{M.I.}},
\bauthor{\bsnm{Wulandari}, \binits{H.R.T.}}:
\bjtitle{AIP Conference Proceedings}
\bvolume{1589},
\bfpage{45}
(\byear{2014})
\end{barticle}
\endbibitem

\bibitem[\protect\citeauthoryear{S\'anchez et~al.}{2010}]{sanchez09}
\begin{barticle}
\bauthor{\bsnm{S\'anchez}, \binits{N.}},
\bauthor{\bsnm{Vicente}, \binits{B.}},
\bauthor{\bsnm{Alfaro}, \binits{E.J.}}:
\bjtitle{\aap}
\bvolume{510A},
\bfpage{78}
(\byear{2010})
\end{barticle}
\endbibitem

\bibitem[\protect\citeauthoryear{Sanders}{1971}]{sanders71}
\begin{barticle}
\bauthor{\bsnm{Sanders}, \binits{W.L.}}:
\bjtitle{\aap}
\bvolume{14},
\bfpage{226}
(\byear{1971})
\end{barticle}
\endbibitem

\bibitem[\protect\citeauthoryear{Silverman}{1986}]{silverman86}
\begin{bbook}
\bauthor{\bsnm{Silverman}, \binits{B.W.}}:
\bbtitle{Density Estimation for Statistics and Data Analysis}.
\bpublisher{{Chapman} \& {Hall}},
\blocation{London}
(\byear{1986})
\end{bbook}
\endbibitem

\bibitem[\protect\citeauthoryear{Vasilevkis et~al.}{1958}]{vasilevkis58}
\begin{barticle}
\bauthor{\bsnm{Vasilevkis}, \binits{S.}},
\bauthor{\bsnm{Klemola}, \binits{A.}},
\bauthor{\bsnm{Preston}, \binits{G.}}:
\bjtitle{\aj}
\bvolume{63},
\bfpage{387}
(\byear{1958})
\end{barticle}
\endbibitem

\bibitem[\protect\citeauthoryear{Wand}{1994}]{wand94}
\begin{barticle}
\bauthor{\bsnm{Wand}, \binits{M.P.}}:
\bjtitle{\jcgs}
\bvolume{3},
\bfpage{433}
(\byear{1994})
\end{barticle}
\endbibitem

\bibitem[\protect\citeauthoryear{Wiramihadja et~al.}{2009}]{wira06}
\begin{barticle}
\bauthor{\bsnm{Wiramihadja}, \binits{S.D.}},
\bauthor{\bsnm{Arifyanto}, \binits{M.I.}},
\bauthor{\bsnm{Sugianto}, \binits{Y.}}:
\bjtitle{\apss}
\bvolume{319},
\bfpage{125}
(\byear{2009})
\end{barticle}
\endbibitem

\bibitem[\protect\citeauthoryear{Zacharias et~al.}{2004}]{zacharias04}
\begin{barticle}
\bauthor{\bsnm{Zacharias}, \binits{N.}},
\bauthor{\bsnm{Urban}, \binits{S.E.}},
\bauthor{\bsnm{Zacharias}, \binits{M.I.}},
\bauthor{\bsnm{Wycoff}, \binits{G.L.}},
\bauthor{\bsnm{Hall}, \binits{D.M.}},
\bauthor{\bsnm{Monet}, \binits{D.G.}},
\bauthor{\bsnm{Rafferty}, \binits{T.J.}}:
\bjtitle{\aj}
\bvolume{127},
\bfpage{3043}
(\byear{2004})
\end{barticle}
\endbibitem

\bibitem[\protect\citeauthoryear{Zacharias et~al.}{2013}]{zacharias13}
\begin{barticle}
\bauthor{\bsnm{Zacharias}, \binits{N.}},
\bauthor{\bsnm{Finch}, \binits{C.T.}},
\bauthor{\bsnm{Girard}, \binits{T.M.}},
\bauthor{\bsnm{Henden}, \binits{A.}},
\bauthor{\bsnm{Bartlett}, \binits{J.L.}},
\bauthor{\bsnm{Monet}, \binits{D.G.}},
\bauthor{\bsnm{Zacharias}, \binits{M.I.}}:
\bjtitle{\aj}
\bvolume{145},
\bfpage{44}
(\byear{2013})
\end{barticle}
\endbibitem

\bibitem[\protect\citeauthoryear{Zhao and He}{1990}]{zhao90}
\begin{barticle}
\bauthor{\bsnm{Zhao}, \binits{J.L.}},
\bauthor{\bsnm{He}, \binits{Y.P.}}:
\bjtitle{\aap}
\bvolume{237},
\bfpage{54}
(\byear{1990})
\end{barticle}
\endbibitem

\bibitem[\protect\citeauthoryear{Zhao et~al.}{2006}]{zhao2006}
\begin{barticle}
\bauthor{\bsnm{Zhao}, \binits{J.L.}},
\bauthor{\bsnm{Chen}, \binits{L.}},
\bauthor{\bsnm{Wen}, \binits{W.}}:
\bjtitle{Chin. J. Astron. Astrophys.}
\bvolume{6},
\bfpage{435}
(\byear{2006})
\end{barticle}
\endbibitem

\end{thebibliography}
